\documentclass[english,british]{elsarticle}
\usepackage[T1]{fontenc}
\usepackage[latin9]{inputenc}
\usepackage{amsmath}
\usepackage{amssymb}
\usepackage{wasysym}
\usepackage{graphicx}
\usepackage{babel}
\begin{document}

\title{\textbf{A Radio Frequency  Helical Deflector for keV Electrons}}

\begin{frontmatter}{}

\author[ye]{L. Gevorgian}

\author[ye]{R. Ajvazyan}

\author[ye]{V. Kakoyan}

\author[ye]{A. Margaryan\corref{cauth}}

\ead{mat@mail.yerphi.am}

\author[gla]{J.R.M. Annand}

\cortext[cauth]{Corresponding author: A.I. Alikhanyan National Science Laboratory
(Yerevan Physics Institute) 2 Br. Alikhanyan Str., 0036 Yerevan, Armenia.
Tel. +374 10 341 500, Fax. +374 10 349 392 }

\address[ye]{A.I. Alikhanyan National Science Laboratory, Yerevan, Armenia}

\address[gla]{Department of Physics \& Astronomy, University of Glasgow, G12 8QQ,
Scotland, UK}

\begin{abstract}
This paper describes a helical deflector to perform circular sweeps
of keV electrons by means of radio frequency fields in a frequency
range 500-1000 MHz. By converting the time dependence of incident
electrons to a hit position dependence on a circle, this device can
potentially achieve extremely precise timing. The system can be adjusted
to the velocity of the electrons to exclude the reduction of deflection
sensitivity due to finite transit time effects. The deflection electrodes
form a resonant circuit, with quality factor Q in excess of 100, and
at resonance the sensitivity of the deflection system is around 1~mm
per V of applied RF input. 

Keywords: Radio Frequency Electron Deflector
\end{abstract}

\end{frontmatter}{}

\section{Introduction }

Circular scanning is commonly used in high-resolution timing devices
such as streak cameras and the first experiments in this technique
were started in the 1950's. Much effort has gone into the development
of deflection systems, which would give a reasonable deflection of
a beam of electrons using moderate RF voltages applied to the deflectors.
Different types of deflectors have been proposed, developed and applied.
Rudenberg \cite{key-1} showed that loss in dynamic sensitivity, due
to transit time effects in parallel plate deflectors, could be reduced
by substituting parallel wire deflectors. The analysing system consisted
of two crossed pairs of shorted Lecher-wires, resonant at the frequency
(3000~MHz) of the beam modulating system under investigation \cite{key-2}.
It was used to measure ps electron bunches \cite{key-3}, where the
time dispersion of secondary electron emission process was determined
to be less than 6 ps \cite{key-4}. 

However analysing systems based on parallel plates have also been
developed \cite{key-5} and applied \cite{key-6,key-7}. In these
circular-scan systems deflection frequencies up to 300~MHz were employed.
Later, circular-scan streak tubes operating at frequencies of 200~MHz
\cite{key-8} and 320~MHz \cite{key-9} have been developed for use
in a laser ranging systems. By applying a 320~MHz, 13~W RF signal,
which had a phase shift applied between the deflection plates, a 6~mm
circular sweep was obtained \cite{key-9}. A circular sweep speed
(which essentially defines the achievable time resolution of the device)
of $6.44\times10^{8}$~cm/s was achieved, providing a temporal resolution
of 30-35~ps and a range difference jitter of less then 6~ps. With
further development, parallel-plate deflection systems have achieved
even faster circular sweeps. A speed of $2.85\times10^{9}$~cm/sec
was achieved by Sibbet \emph{et al.} by applying 300~MHz, 15~W RF
power to a Photochron IIC streak tube \cite{key-10}. A 30~mm diameter
scan was obtained, enabling an instrumental time resolution better
then 6~ps. $6.25\times10^{9}$~cm/s was achieved by Bryukhnevitch
\emph{et al}. by applying 500~MHz, 17~W RF power to a PV006S streak
tube \cite{key-11}. A 40~mm diameter circular scan and time resolution
of less than 5~ps were achieved. 

At higher frequencies other types of deflection systems have been
investigated in the 1970's. A travelling wave RF analysing system
was used by Kalibjian \emph{et al.} \cite{key-12}, giving a time
resolution of about 0.7~ps when a 3.6~GHz RF field was applied.
Alternatively a 10~GHz slit-aperture-resonator deflection system
was employed by Butslov \emph{et al}., giving a time resolution of
0.5~ps \cite{key-13} and a similar RF deflector has been investigated
by the Yerevan group for 3~GHz frequency applications \cite{key-14}.
More recently a cylindrical RF cavity was used to perform circular
sweeps, using a transverse magnetic field in TM110 mode, with circular
polarization and a frequency of 2.856 GHz \cite{key-15}. The time
resolution was about 0.6 ps.

In this article we present a sensitive and compact circular sweep
deflection system for keV electrons, capable of operating in the 500-1000~MHz
frequency range. The theory of the deflector and calculations based
on simulation studies are described in Sec.\ref{sec:The-Ultra-High-Frequency}.
Results of experimental studies of the deflector are presented in
Sec.\ref{sec:Experimental-studies}.

\selectlanguage{english}%

\section{\label{sec:The-Ultra-High-Frequency}\foreignlanguage{british}{The
Circular Scanning Deflector For keV Electrons }}

\selectlanguage{british}%
When an electron beam passes through a parallel-plate deflection system
that has an axial length $\Lambda$, a separation $d$ and a potential
difference $U_{d}$ between the plates (Fig. 1), it is deflected vertically
by an angle $\tan\theta=\dfrac{eU_{d}\varLambda}{mdv_{z}^{2}}$, where
$e,m,v_{z}$ are respectively the charge, mass and velocity of the
electron. The accelerating potential $U_{a}$ of the electron ``gun''
of the system produces an energy $eU_{a}=\dfrac{mv_{z}^{2}}{2}$ and
hence $\tan\theta=\dfrac{U_{d}\varLambda}{2U_{a}d}$. This produces
a deflection $\Delta Y$ at a distance $D$ from the plates:

\begin{equation}
\Delta Y=D\tan\theta=\dfrac{DU_{d}\varLambda}{2U_{a}d}\label{eq:DeltaY}
\end{equation}

The static sensitivity of the $Y$ deflection, i.e. the sensitivity
of the deflector at low frequencies, is defined by $\varepsilon_{s}=\Delta Y/U_{d}$.
The main problem for transverse deflection of keV electrons at high
frequencies is connected to the finite time that electrons take to
travel through the deflector. If the RF field changes according to
$\sin\omega t$, where $\omega=2\pi/T$ and $T$ is the period of
the RF Voltage, the dynamic sensitivity of the parallel plate RF deflector
$\varepsilon_{d}$, can be expressed as\foreignlanguage{english}{
\begin{equation}
\varepsilon_{d}=\varepsilon_{s}\dfrac{\sin\left(\omega T_{0}/2\right)}{\left(\omega T_{0}/2\right)}\label{eq:Eps_d}
\end{equation}
}\\
where $T_{0}=\Lambda/v_{z}$ is the transit time of the electron through
the deflector and $\omega T_{0}$ is the so called transit angle.
The dynamic sensitivity is a function of $\omega,\Lambda,v_{z}$ and
at very high frequencies the transit time of the electrons through
the deflecting field reduces the magnitude of the deflection angle,
i.e. $\varepsilon_{d}<\varepsilon_{s}$ unless the transit time is
smaller than a half-period of the highest frequency to be used (see
Ref.\cite{key-1,key-16}). 

\selectlanguage{english}%
\begin{figure}
\begin{center}\includegraphics[width=0.6\columnwidth]{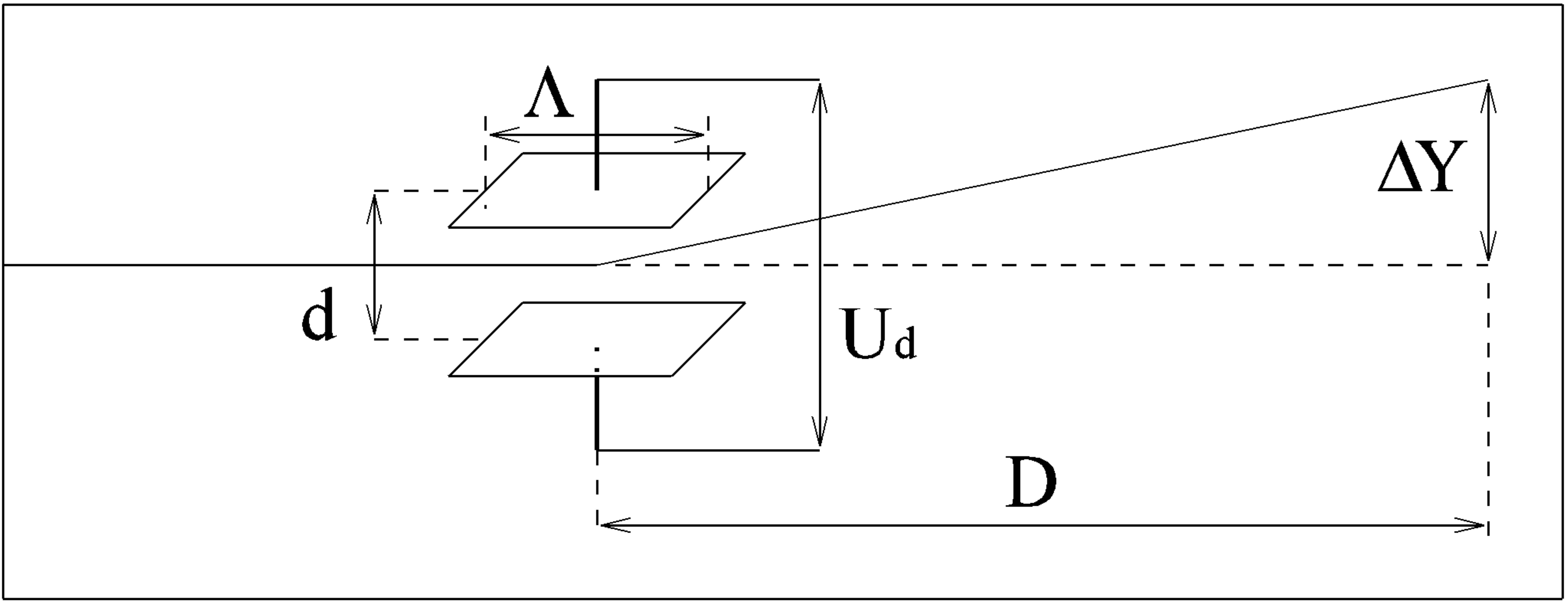}\end{center}

\protect\caption{\label{fig:Parallel-plate}\foreignlanguage{british}{Schematic of
a parallel-plate deflection system.}}
\end{figure}

\begin{figure}
\begin{center}\includegraphics[width=0.7\columnwidth]{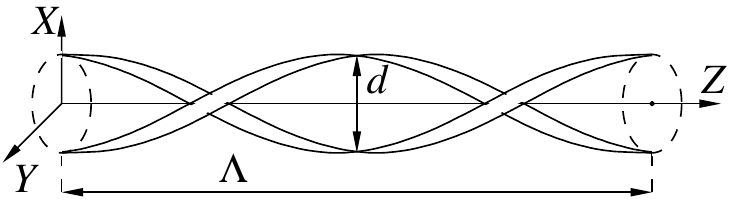}\end{center}

\selectlanguage{british}%
\protect\caption{\selectlanguage{english}%
\label{fig:Shamaev}Schematic of the Shamaev helical RF deflector.\selectlanguage{british}%
}
\selectlanguage{english}%
\end{figure}

\selectlanguage{british}%
To avoid transit time effects at high frequencies a dedicated RF deflector
was proposed by Shamaev%
\footnote{The Ph.D. thesis of Y.M. Shamaev is referred to in Ref.\cite{key-17},
but we were unable to access this unpublished document.%
} in the 1960's (the theory is described in Ref.\cite{key-16}). The
Shamaev deflector is based on a pair of helical electrodes (Fig. \ref{fig:Shamaev})
of periodic length $\Lambda$ and separation $d$. Electrons move
in the direction of the Z axis, which coincides with the axis of the
deflection system, with a constant velocity $v_{z}$. An applied sinusoidal
RF Voltage across the deflection plates produces a transverse ($E_{z}=0$)
electric field so that the equation of motion of electrons in this
field can be written as: $\dfrac{dv_{\perp}}{dt}=-\dfrac{eE_{\perp}}{m}$,
$\dfrac{dv_{z}}{dt}=0$, where $v_{\perp}=v_{x}+iv_{y}$ and $E_{\perp}=E_{x}+iE_{y}$
are complex variables.

\selectlanguage{english}%

\selectlanguage{british}%
When the wavelength of the applied RF field $\lambda\gg\Lambda$,
one can to first approximation ignore the variation of the field along
the length of the electrode, in which case $E_{\perp}\thickapprox E\left(t\right)\exp\left(i2\pi z/\Lambda\right)$,
where $E\left(t\right)=E_{d}\sin\left(\omega t+\phi\right)$, $E_{d}=U_{d}/d$
is the amplitude of the electric field in the deflector for an applied
RF voltage of $U_{d}$ and $\phi$ is the RF phase at entry of an
electron into the deflector. The oscillating field can be represented
by a sum of two fields which sweep back and forth, by means of the
Euler formula:

\begin{equation}
\sin\left(\omega t+\phi\right)=\dfrac{1}{2i}\left\{ \exp\left[i\left(\omega t+\phi\right)\right]-\exp\left[-i\left(\omega t+\phi\right)\right]\right\} \label{eq:euler}
\end{equation}
\\
Assuming that electrons with constant velocity $v_{z}$ enter the
RF deflector at $t=0$, the equation of motion in the $X-Y$ plane
is given by:

\begin{equation}
\dfrac{dv_{\perp}}{dt}=\tfrac{eE_{d}}{2im}\left\{ \exp\left[i\left(\omega_{c}-\omega\right)t-i\phi\right]-\exp\left[i\left(\omega_{c}+\omega\right)t+i\phi\right]\right\} \label{eq:motion-XY}
\end{equation}
\\
where $\omega_{c}=\frac{2\pi z}{\Lambda t}=\frac{2\pi v_{z}}{\Lambda}=\frac{2\pi}{T_{c}}$.
The transverse velocity of the electron after the deflector $v_{\perp}$
can be determined by integration of Eq.\ref{eq:motion-XY} over the
range $t=0-\tau$, where $\tau=l/v_{z}$ and $l$ is the length of
the deflector. $T_{c}$ is the transit time over one periodic length
of the deflector ($l=\Lambda$).

{\small{}
\begin{eqnarray}
v_{\perp} & = & \dfrac{eE_{d}}{2m}\left\{ \dfrac{\exp\left[i\left(\omega_{c}+\omega\right)\right]\tau-1}{\omega_{c}+\omega}\exp\left(i\phi\right)-\dfrac{\exp\left[i\left(\omega_{c}-\omega\right)\right]\tau-1}{\omega_{c}-\omega}\exp\left(-i\phi\right)\right\} \label{eq:v-perp}\\
 & = & i\dfrac{eE_{d}\tau}{2m}\left\{ \dfrac{\sin x_{2}}{x_{2}}\exp i\left(x_{2}+\phi\right)-\dfrac{\sin x_{1}}{x_{1}}\exp i\left(x_{1}-\phi\right)\right\} \nonumber 
\end{eqnarray}
}\\
where $x_{1}=\frac{\omega_{c}-\omega}{2}\tau=\frac{\omega_{c}-\omega}{2v_{z}}l$
and $x_{2}=\frac{\omega_{c}+\omega}{2}\tau=\frac{\omega_{c}+\omega}{2v_{z}}l$.
The $x,y$ components of the transverse velocity are:

\begin{eqnarray}
v_{x} & = & -\dfrac{eE_{d}\tau}{2m}\left\{ \dfrac{\sin x_{2}}{x_{2}}\sin\left(x_{2}+\phi\right)-\dfrac{\sin x_{1}}{x_{1}}\sin\left(x_{1}-\phi\right)\right\} \label{eq:v-perp-x}\\
v_{y} & = & -\dfrac{eE_{d}\tau}{2m}\left\{ -\dfrac{\sin x_{2}}{x_{2}}\cos\left(x_{2}+\phi\right)+\dfrac{\sin x_{1}}{x_{1}}\cos\left(x_{1}-\phi\right)\right\} \nonumber 
\end{eqnarray}

\selectlanguage{english}%
\begin{figure}
\selectlanguage{british}%
\includegraphics[width=1\columnwidth]{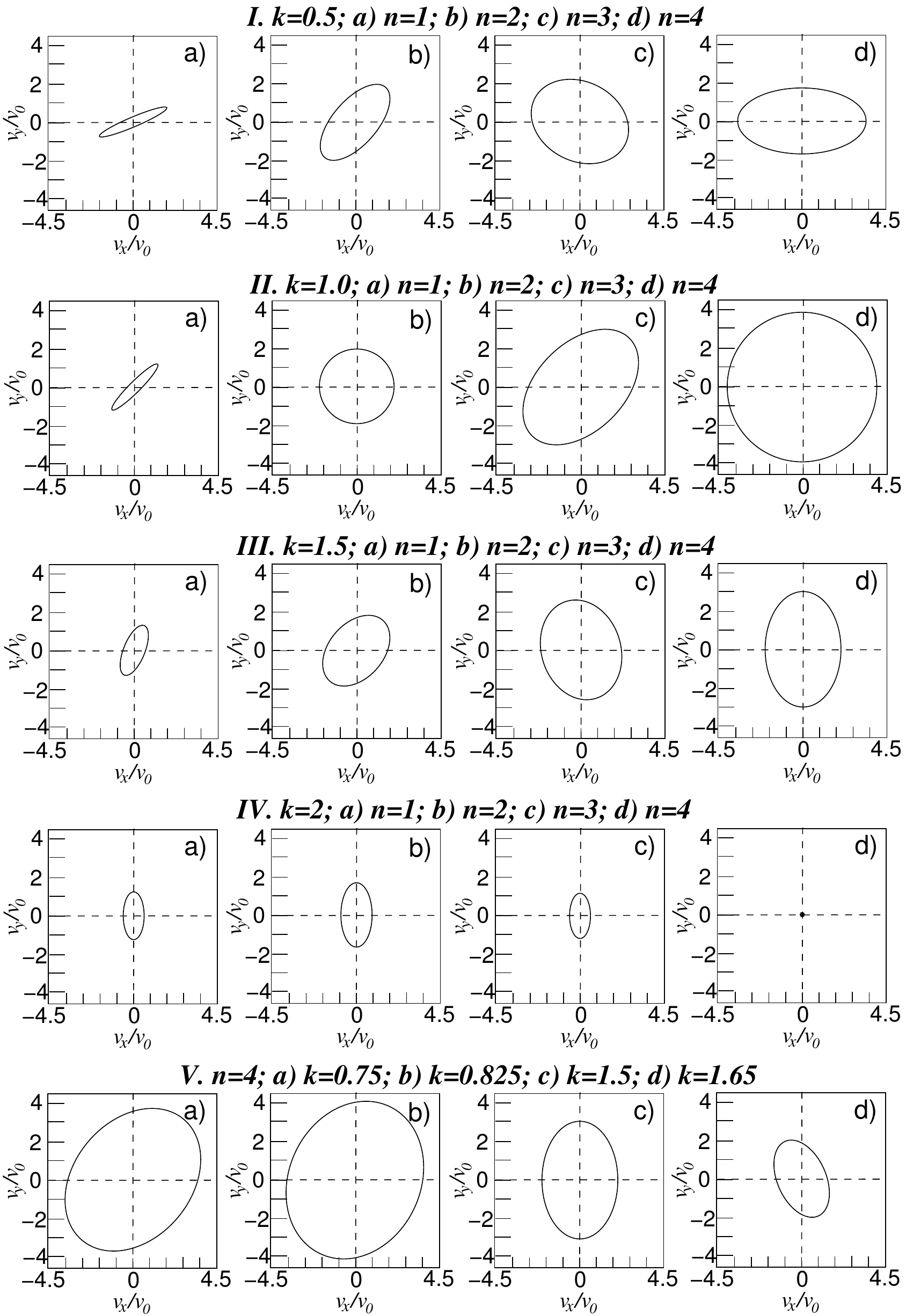}

\selectlanguage{english}%
\protect\caption{\selectlanguage{british}%
\label{fig:Transverse-velocity}Transverse velocity components $v_{x}$
vs. $v_{y}$ in units of $v_{0}$.\selectlanguage{english}%
}

\selectlanguage{british}%
\selectlanguage{english}%
\end{figure}

\selectlanguage{british}%
Simulations were carried out to investigate the behaviour of the helical
RF deflector at different values of $k=\omega/\omega_{c}$ and $l=n\Lambda/4$
($n=1,2,3,4)$. Plots of $v_{x}$ vs. $v_{y}$ in units of $v_{0}=eE_{d}\Lambda/8mv_{z}$,
for different values of $k$ and $n$, are presented in Fig.3. These
studies show that the $x-y$ locus is circular only for the $k=1$
and $n=2,4$ cases. In the following we concentrate on the case $k=1,$
$n=4$ where the transit time of electrons through the deflector with
$l=\Lambda$ is equal to the period of the applied high frequency
sinusoidal voltage ($\omega=\omega_{c}$). We refer to this as the
Shamaev resonance condition, where one has maximum transverse velocity
$v_{\perp}=4v_{0}$, and Eq.\ref{eq:v-perp-x}, takes a simple form.
In the case of $k=1,$ $n=4$ then $x_{1}=0,$ $x_{2}=2\pi$, leading
to $\sin x_{1}/x_{1}=1,$ $\sin x_{2}=0$ and thus $v_{x}=-4v_{0}\sin\phi,$
$v_{y}=-4v_{0}\cos\phi$. The deflection angle is given by:

\selectlanguage{english}%
\begin{equation}
\tan\theta=\dfrac{v_{\perp}}{v_{z}}=\dfrac{eE_{d}\tau}{2mv_{z}}\sqrt{\left(\dfrac{\sin x_{1}}{x_{1}}-\dfrac{\sin x_{2}}{x_{2}}\right)^{2}+4\dfrac{\sin x_{1}\sin x_{2}}{x_{1}x_{2}}\sin^{2}\left(\dfrac{\omega\tau}{2}+\phi\right)}\label{eq:defl-angle}
\end{equation}
\\
\foreignlanguage{british}{which for $k=1$ and $n=4$ reduces to }$\tan\theta=\frac{4v_{0}}{v_{z}}=\frac{eE_{d}\Lambda}{2mv_{z}^{2}}=\frac{U_{d}\Lambda}{4dU_{a}}$.
\foreignlanguage{british}{When the Shamaev resonance condition is
applied to a helical deflector, transit-time effects are avoided and
$\varepsilon_{d}=\varepsilon_{s}/2$ (the factor 2 in the denominator
is related to the helical shape of the deflector electrodes). Electron
trajectories can be analysed in two parts: those inside and outside
the deflector. The electron's transverse coordinates at the end of
the RF deflector, $r_{x}=\varint v_{x}dt$ and $r_{y}=\varint v_{y}dt$,
are obtained by integrating over the range $t=0-l/v_{z}$, resulting}
\foreignlanguage{british}{in:}

{\small{}
\begin{eqnarray}
r_{x} & = & \dfrac{eE_{d}l^{2}}{4mv_{z}^{2}}\left\{ \cos\phi\left[\dfrac{1}{x_{1}}\left(1-\dfrac{\sin2x_{1}}{2x_{1}}\right)+\dfrac{1}{x_{2}}\left(1-\dfrac{\sin2x_{2}}{x_{2}}\right)\right]+\sin\phi\left[\dfrac{\sin^{2}x_{1}}{x_{1}^{2}}+\dfrac{\sin^{2}x_{2}}{x_{2}^{2}}\right]\right\} \label{eq:rxy}\\
r_{y} & = & \dfrac{eE_{d}l^{2}}{4mv_{z}^{2}}\left\{ -\sin\phi\left[\dfrac{1}{x_{1}}\left(1-\dfrac{\sin2x_{1}}{2x_{1}}\right)-\dfrac{1}{x_{2}}\left(1-\dfrac{\sin2x_{2}}{x_{2}}\right)\right]+\cos\phi\left[\dfrac{\sin^{2}x_{1}}{x_{1}^{2}}+\dfrac{\sin^{2}x_{2}}{x_{2}^{2}}\right]\right\} \nonumber 
\end{eqnarray}
}\\
\foreignlanguage{british}{For $k=1$ and $n=4$ these formulae simplify
to $r_{x}=R_{0}\left[\sin\phi+\cos\phi/2\pi\right]$ and $r_{y}=R_{0}\left[\cos\phi+\sin\phi/2\pi\right]$,
where $R_{0}=\frac{eE_{d}\Lambda^{2}}{4mv_{z}^{2}}=\frac{U_{d}\Lambda^{2}}{8dU_{a}}$.
At the end of the deflector the electrons will scan through an ellipse
of half axes $a=R_{0}\left[1+1/2\pi\right]$ and $b=R_{0}\left[1-1/2\pi\right]$.
If the electrons are detected on a screen at a distance $D$ from
the deflector, the transverse coordinates will be $R_{x}=r_{x}+D\tan\theta_{x}$
and $R_{y}=r_{y}+D\tan\theta_{y}$, where $\tan\theta_{x}=-4v_{0}\sin\phi/v_{z}$
and $\tan\theta_{y}=-4v_{0}\cos\phi/v_{z}$, leading to: }

\selectlanguage{british}%
\begin{eqnarray}
R_{x} & = & r_{x}-D\dfrac{4v_{0}\sin\phi}{v_{z}}\label{eq:Rxy}\\
R_{y} & = & r_{y}-D\dfrac{4v_{0}\cos\phi}{v_{z}}\nonumber 
\end{eqnarray}
\\
 For $U_{a}=2.5$~kV and $\Lambda=6\:(3)$~cm, corresponding to
Shamaev resonance conditions at 500 (1000)~MHz, one obtains, for
$U_{d}=20$~V and $d=1$~cm, $R_{0}=0.036$ (0.009)~cm and $\tan\theta=0.012$
(0.006). At distances $D=0,\:12,\:24$~cm, circles of radius 1, 5,
9 (1, 9, 17) $R_{0}$ respectively are produced. In Sec.\ref{sec:Experimental-studies}
it is demonstrated that the deflector forms a resonant circuit with
a Q-factor in excess of 100 at these frequencies. This results in
an observed deflection more than an order of magnitude larger than
has been calculated here.

\selectlanguage{english}%

\section{\label{sec:Experimental-studies}Experimental studies }

\begin{figure}
\selectlanguage{british}%
\begin{center}\includegraphics[width=0.8\columnwidth]{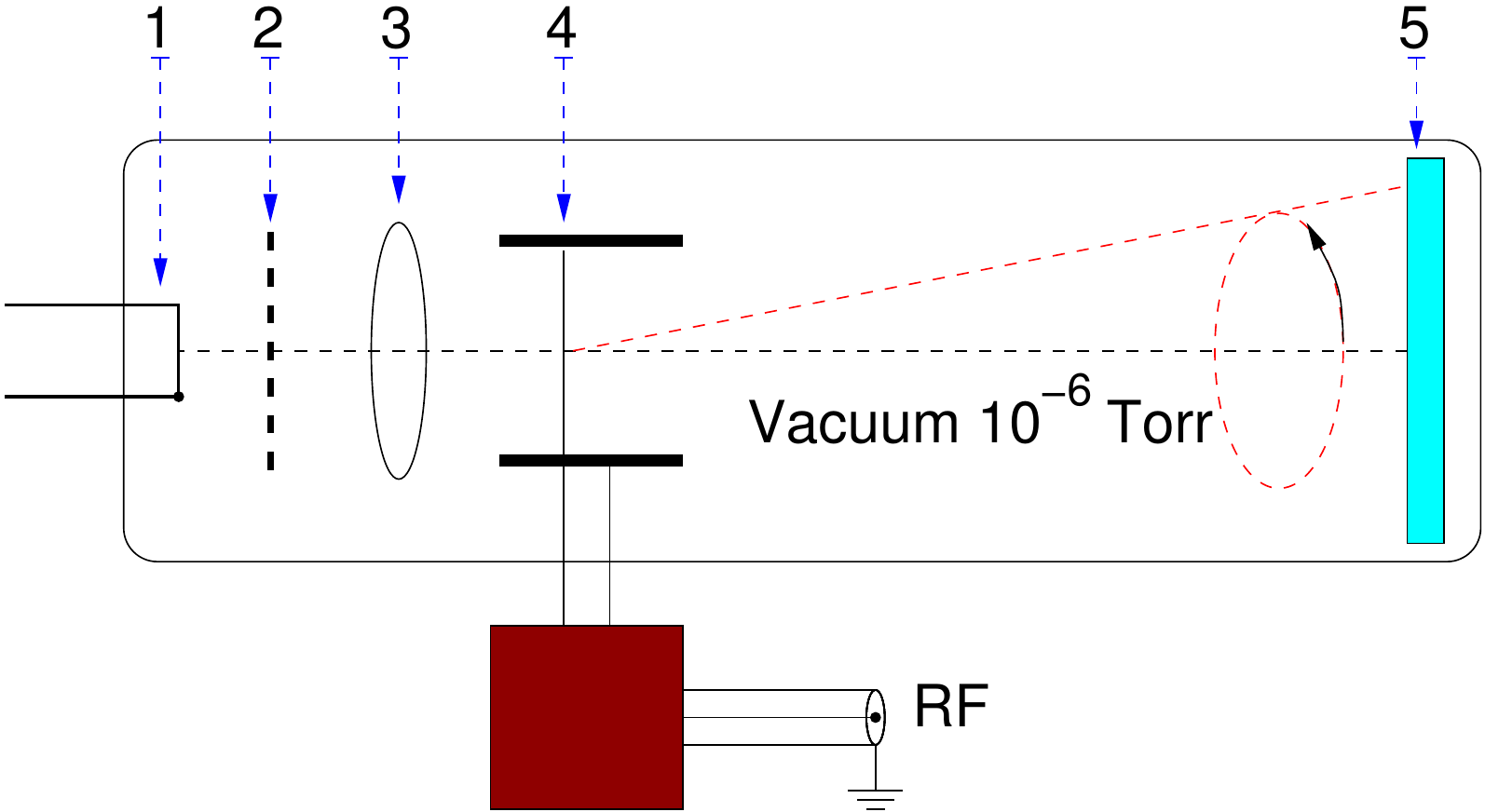}\end{center}

\selectlanguage{english}%
\protect\caption{\selectlanguage{british}%
\label{fig:Exp-schematic}A schematic representation of the experimental
set up: 1 thermionic cathode, 2 electron transparent accelerating
electrode, 3 electrostatic lens, 4 electrodes of the RF deflector,
5 phosphor screen.\selectlanguage{english}%
}
\end{figure}

\selectlanguage{british}%
The experimental set up for investigation of the characteristics of
the RF deflector is presented schematically in Fig.\ref{fig:Exp-schematic}.
It consists of a pumped vacuum vessel with various electrical and
mechanical feed throughs. A thermionic cathode (1) emits electrons,
which are accelerated by a voltage applied between the cathode and
an electron transparent electrode (2). An electrostatic lens (3) then
focuses the accelerated electrons on to the phosphor screen (5) at
the far end of the tube. The RF deflection system, consists of electrodes
(4), a RF source and tuning circuit. The experimental commissioning
followed the procedure: i. mechanical and vacuum testing; ii. electron
beam optics tuning; iii. RF deflector tuning.

\begin{figure}
\includegraphics[width=1\columnwidth]{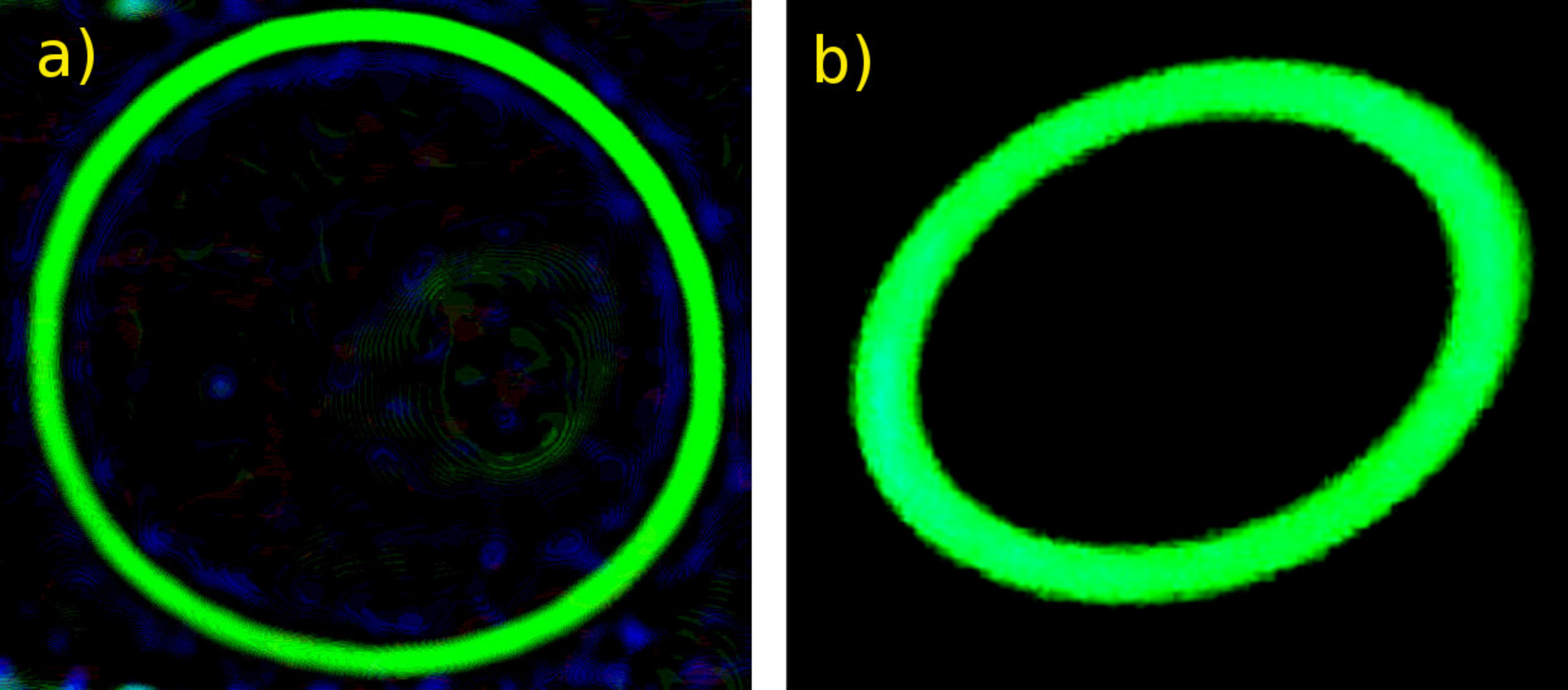}

\protect\caption{\label{fig:Images-deflected}Images of the deflected 2.5 keV continuous
electron beam at the phosphor screen. The parameters of the deflector
and test set up are: $\Lambda=3$~cm, $d=1$~cm, $U_{d}=20$~V,
$U_{a}=2.5$~kV, $D=12$~cm. Applied RF frequencies are: a) 1000
MHz; b) 750 MHz. The radius of the scanned circle is 2 cm. The length
of the major axis of the ellipse is also around 2~cm.}
\end{figure}

The size of the focused thermionic electron beam spot on the phosphor
screen, mounted 12~cm from the end of the deflector, was less than
1~mm diameter. The RF deflection system consisted of one pair of
helical electrodes (Fig. \ref{fig:Shamaev}), which served as a capacitive
element, containing the applied RF power within the deflector. For
a given set of electrodes we observed that at some frequencies the
deflection angles, or radii of the scanned circles, were about an
order magnitude higher than anticipated from theory. Indeed for $U_{d}=20$~V,
$\omega=1000$~MHz, $\Lambda=3$~cm, $d=1$~cm and $U_{a}=2.5\, k$V,
the radius of the scanning circle at a distance $D=12$~cm was predicted
to be 0.81~mm, while it was measured at 20~mm (Fig. \ref{fig:Images-deflected}a).
For a tuned deflector at $\omega=750$~MHz frequency, electrons scan
on to an ellipse (Fig. \ref{fig:Images-deflected}b). These two cases
correspond to the theoretical predictions presented in Fig.\ref{fig:Transverse-velocity}-IId
(n = 4, k = 1) and Fig.\ref{fig:Transverse-velocity}-Va (n = 4, k
= 0.75). The observed factor $\sim25$ larger deflection than predicted
by Eq.\ref{eq:Rxy} can be explained if the deflection electrodes
form a resonant circuit at these frequencies with a high Q-factor.
The resonant frequency can be fixed at a desired value, i.e. near
the Shamaev resonant frequency, by tuning the capacitance of the deflection
system. This may be realized by suitable design of the electrodes
or, for given electrodes, by tuning of the deflector capacitance using
a coaxial cavity or an additional variable capacitor. 

\begin{figure}
\begin{center}\includegraphics[width=0.6\columnwidth]{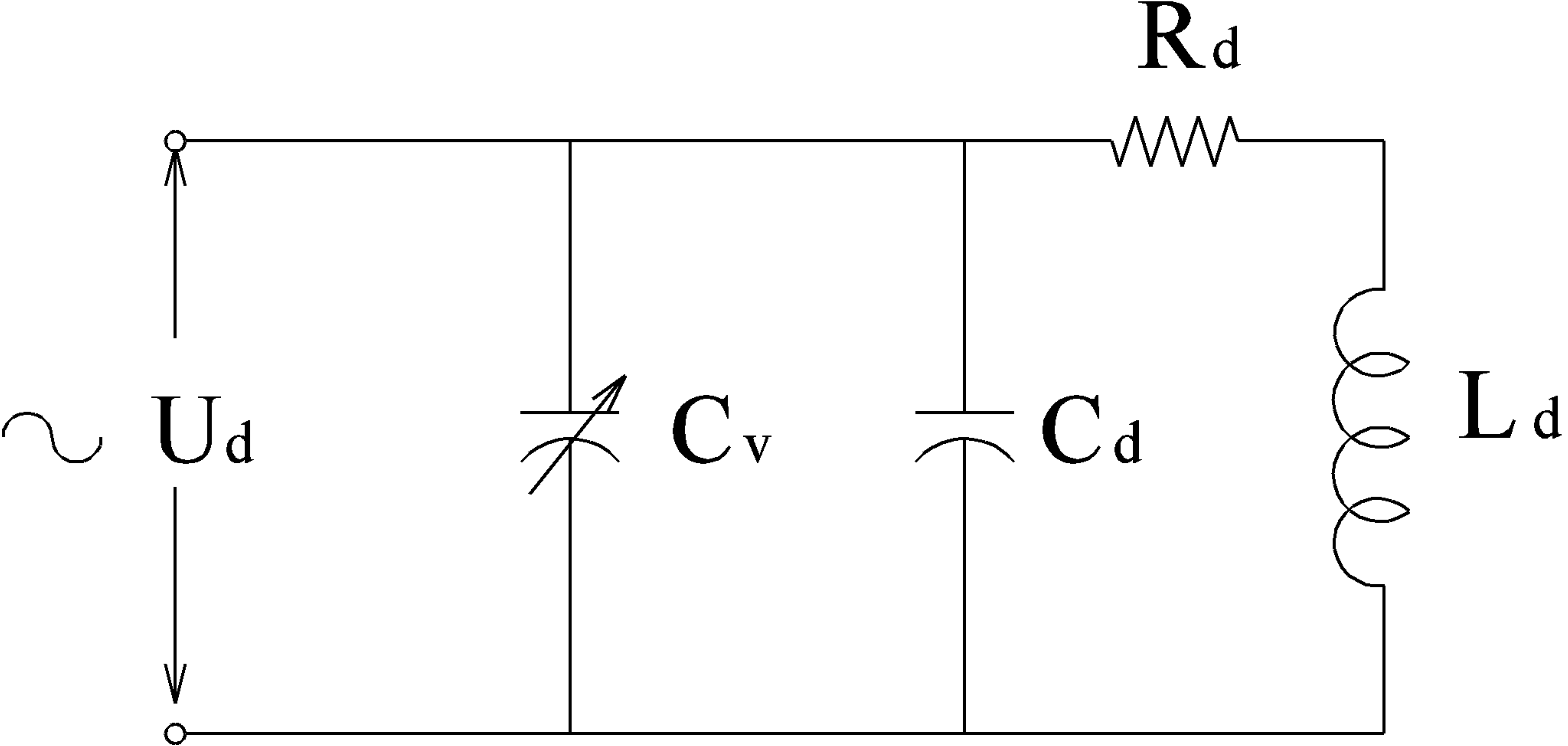}\end{center}

\protect\caption{\label{fig:equivalent-circuit}An equivalent circuit of the RF deflector
and resonant tuning system. $U_{d}$: applied RF Voltage; $C_{v}$:
variable capacitance; $C_{d}$: deflector capacitance; $R_{d}$ deflector
resistance; $L_{d}$: deflector inductance. }
\end{figure}

\begin{figure}
\begin{center}\includegraphics[width=0.7\columnwidth]{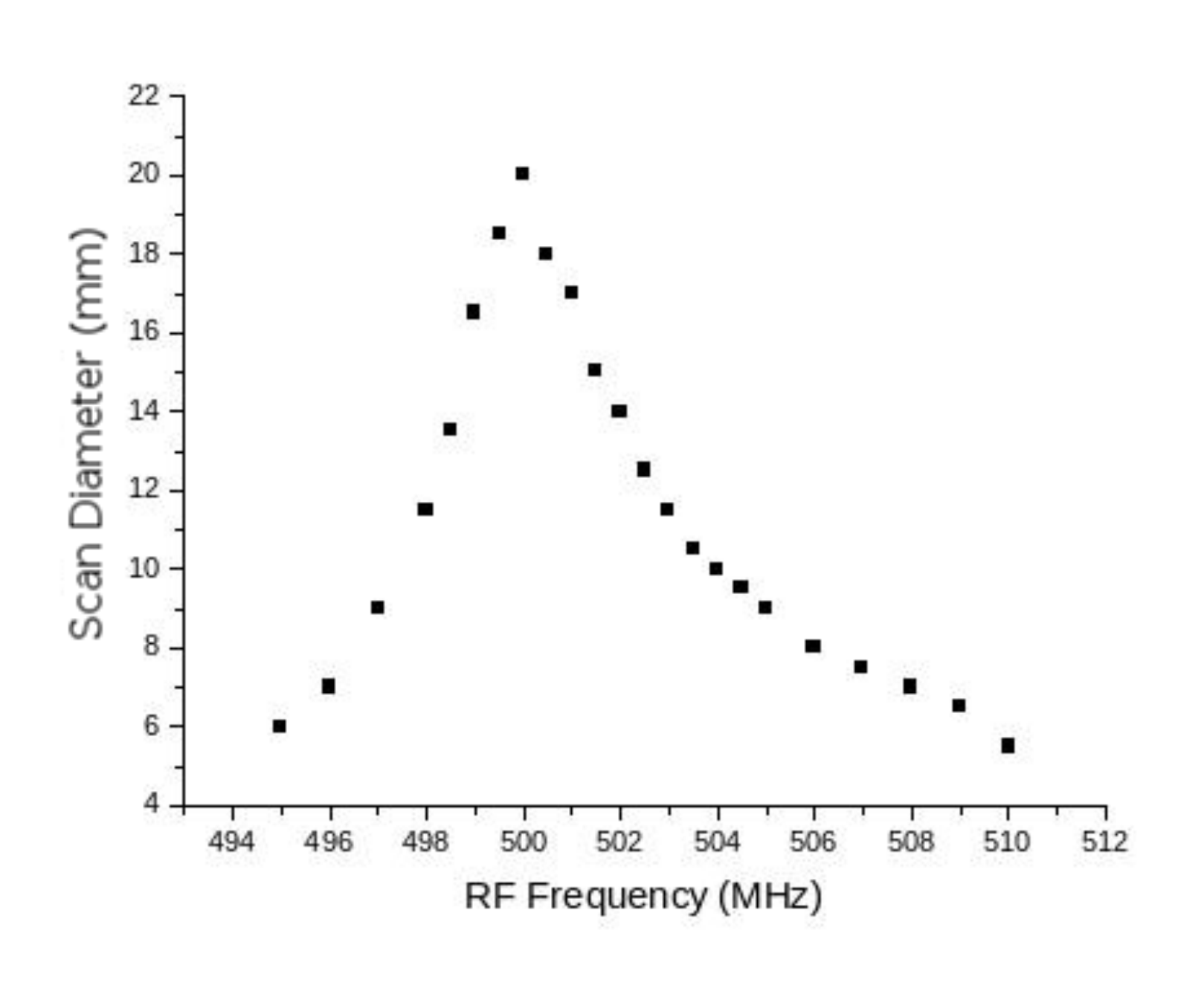}\end{center}

\protect\caption{\label{fig:DiameterQ}Diameter of the scanning circle as a function
of RF frequency for the 500 MHz helical deflector. The parameters
of the deflector and test set up are: $\Lambda=6$~cm, $d=1$~cm,
$U_{d}=10$~V, $U_{a}=2.5$~kV, $D=12$~cm.}
\end{figure}

The equivalent circuit of the helical deflector and tuning system
is presented in Fig.\ref{fig:equivalent-circuit}. The Q-factor of
the resonant circuit was determined experimentally, by measuring the
diameter of the scanning circle as a function of the applied RF frequency.
An example of the resonant behaviour at frequencies around 500 MHz
for a deflector with parameters $\Lambda=6$~cm, $d=1$~cm, $U_{d}=10$~V,
$U_{a}=2.5$~kV, $D=12$~cm. is presented in Fig.\ref{fig:DiameterQ}.
From this it follows that the Q-factor of the formed resonance is
about 130. Thanks to this resonance, about 1~W (into $50\Omega$)
of RF power is enough to scan 2.5 keV electrons circularly at a radius
of 2 cm. The sensitivity of this new and compact RF deflector is about
1~mm/V or 0.1~rad/$W^{1/2}$ which is an order of magnitude higher
than the sensitivities of other RF deflectors used previously. Potentially
it has a number of applications in fixed-frequency cathode-ray-tube
based instruments, for example opto-electronic devices such as RF
Streak Cameras or the RF Photomultipler Tube \cite{key-18,key-19}.
The latter has been proposed for ps resolution timing measurements
in nuclear physics experiments.

\selectlanguage{english}%

\section{\label{sec:Summary}Summary }

\selectlanguage{british}%
This paper describes a RF deflector of helical geometry to perform
circular sweeps of keV energy electrons using high frequency electromagnetic
fields. The sweep can be used convert the temporal dependence of a
pulse of electrons passing through the deflector to a position dependence
on a circle. By employing a position sensitive electron detector very
precise timing information at the ps level may then be obtained. A
major advantage of the helical electrodes is that the system can be
optimised to the velocity of the transiting electrons, so that loss
of deflection sensitivity due to finite transit time effects is avoided.

The theory of operation of the deflector is derived and optimum deflector
dimensions calculated for operational frequencies of 500 and 1000~MHz.
After fine tuning the effective capacitance of the system, the deflection
electrodes form a resonant circuit, with a quality factor Q in excess
of 100. On resonance, the sensitivity of the deflection system is
about 1~mm/V or 0.1~rad/$W^{1/2}$and around 1~W (into $50\Omega$)
of RF power is sufficient to scan 2.5 keV electrons circularly at
a radius of 2 cm radius. This makes the deflector suitable for ultra-high
resolution timing devices such as the Radio Frequency Photomultiplier
Tube \cite{key-18,key-19}, where development continues. Future developments
include a dual helical deflection system to produce a spiral rather
than circular scanning pattern.

\paragraph{Acknowledgements}

This work is supported in part by Grant Number 9-8/AB of the State
Committee of Science, Republic of Armenia and by the UK Science and
Technology Facilities Council (STFC 57071/1, 50727/1).

\end{document}